# Broadband waveguide quantum memory for entangled photons

Erhan Saglamyurek[1*], Neil Sinclair[1*], Jeongwan Jin[1*], Joshua A. Slater[1*], Daniel Oblak[1], Félix Bussières[1†], Mathew George[2], Raimund Ricken[2], Wolfgang Sohler[2] & Wolfgang Tittel[1]

[1]*Institute for Quantum Information Science, and Department of Physics & Astronomy, University of Calgary, 2500 University Drive NW, Calgary, Alberta T2N 1N4, Canada*

[2]*Department of Physics - Applied Physics, University of Paderborn, Warburger Str. 100, 33095 Paderborn, Germany*

**The reversible transfer of quantum states of light in and out of matter constitutes an important building block for future applications of quantum communication: it allows synchronizing quantum information[1], and enables one to build quantum repeaters[2] and quantum networks[3]. Much effort has been devoted worldwide over the past years to develop memories suitable for the storage of quantum states[1]. Of central importance to this task is the preservation of entanglement, a quantum mechanical phenomenon whose counter-intuitive properties have occupied philosophers, physicists and computer scientists since the early days of quantum physics[4]. Here we report, for the first time, the reversible transfer of photon-photon entanglement into entanglement between a photon and collective atomic excitation in a solid-state device. Towards this end, we employ a thulium-doped lithium niobate waveguide in conjunction with a photon-echo quantum memory protocol[5], and increase the spectral acceptance from the current maximum of 100 MHz[6] to 5 GHz. The entanglement-preserving nature of our storage device is assessed by comparing the amount of entanglement contained in the detected photon pairs before and after the reversible transfer, showing, within statistical error, a perfect mapping process. Our integrated, broadband quantum memory complements the family of robust, integrated lithium niobate devices[7]. It renders frequency matching of light with matter interfaces in advanced applications of quantum communication trivial and institutes several key properties in the quest to unleash the full potential of quantum communication.**

Quantum communication is founded on the encoding of information, generally referred to as quantum information, into quantum states of light[8]. The resulting applications of quantum physics at its fundamental level offer cryptographic security through quantum key distribution without relying on unproven mathematical assumptions[9] and allow for the disembodied transfer of quantum states between distant places by means of quantum teleportation[8,10]. Reversible mapping of quantum states between light and matter is central to advanced applications of quantum communication, e.g. quantum repeaters[2] and quantum networks[3], in which matter

---

* These authors contributed equally to this work.
† Present address: GAP-Optique, University of Geneva, Rue de l'École-de-Médecine 20, 1211 Geneva 4, Switzerland

constitutes nodes that hold quantum information until needed, and thereby synchronize the information flow through the communication channel or network. Furthermore, such a quantum interface allows the generation of light-matter entanglement through the mapping of one of two entangled photons into matter. Scrutinizing if, and how, different physical system can be entangled, and localizing the fundamental or technological boundaries where this fascinating quantum link breaks down is of central importance to the understanding of quantum physics and has received much attention over the past decades[4,8,10].

The reversible light-matter interface can be realized through the direct transfer of quantum states from light onto matter and back, or through the generation of light-matter entanglement followed by teleportation of quantum information from an externally provided photon into matter, and eventually back. Experimental capabilities have advanced rapidly over the past years and quantum state transfer between light and atomic vapour[11-15], solid state ensembles[16,17], or single absorbers[18], as well as the generation of light-matter entanglement through the emission of photons from atomic ensembles[19-21] or single emitters[22-25] has been reported.

For quantum memory to become practical, it is important to reduce the complexity of experimental implementations and improve their robustness. The recent addition of rare-earth-ion-doped (RE) crystals[16,17] to the set of storage materials has been an important first step towards this goal. Their promise is further enhanced through potentially long storage times - up to several seconds in $Pr:Y_2SiO_5$[26]. In addition, given the large inhomogeneous broadening of optical zero-phonon lines, up to ~100 GHz, RE crystals in principle offer storage of photons with less than 100 ps duration when being used in conjunction with a suitable quantum memory protocol[5]. Yet, the reversible state transfer between light and solid-state devices has so far not been shown to preserve entanglement. This is largely due to the limited spectral bandwidth of current implementations, 100 MHz at most[6], which is orders of magnitude smaller than that of entangled photon pairs generated in the widely employed process of spontaneous parametric down-conversion (SPDC)[8]. In this work, we approach the problem from both ends: we increase the acceptance bandwidth of our storage device to 5 GHz and narrow the bandwidths of our entangled photons to similar values. Furthermore, using a wave-guiding storage medium, we move fundamental quantum memory research further towards application.

The layout of our experiment is depicted in Fig. 1. Short pulses of light travel through an unbalanced interferometer that splits them coherently into *early* and *late temporal modes*. Subsequent SPDC followed by narrow spectral filtering results in the creation of pairs of photons, centred at wavelengths around 795 and 1532 nm, in the *time-bin entangled* qubit state[27]

$$|\phi^+\rangle = \frac{1}{\sqrt{2}}\big(|e,e\rangle + |l,l\rangle\big)_.$$

(1)

Here, $|e\rangle$ and $|l\rangle$ denote *early* and *late* modes and replace the usual *spin-down* and *spin-up* notation for spin-1/2 particles. More specifically, $|i,j\rangle$ denotes a quantum state where the 795 nm photon has been created in the temporal mode $i$, and the 1532 nm photons in $j$. We point out that, due to the filtering, our source generates frequency uncorrelated entangled photons at wavelengths that match the low-loss windows of free-space and standard telecommunication fibre. It can thus readily be employed in real-world applications of quantum communication that involve quantum teleportation and entanglement swapping.

The 1532 nm photon is directed to a qubit analyzer. It consists of either a fibre delay line followed by a single-photon detector (SPD) that monitors the photon's arrival time, or a fibre-optical interferometer that is similarly unbalanced as the pump interferometer. In the latter case, both outputs of the interferometer are connected to SPDs. The role of the delay line followed by detectors is to perform projection measurements of the photon's state onto *early* and *late* qubit states. Alternatively, the interferometer enables projections onto equal superpositions of *early* and *late* modes. Using the language of spin ½ systems, this corresponds to projections onto $\sigma_z$ and, for appropriately chosen phases, $\sigma_x$ and $\sigma_y$, respectively.

The 795 nm photon is transmitted to the quantum memory where its state, specifically that it is entangled with the 1532 nm photon, is mapped onto a collective excitation of millions of thulium ions. After a preset but in principle adjustable time, the state is mapped back onto a photon, which exits the memory through a fibre in well-defined spatio-temporal modes. It is probed by a qubit analyzer that consists of a delay line followed by an SPD, or an interferometer with one output connected to an SPD, similar to what is described above.

To reversibly map the 795 nm photon onto matter, we employ a photon-echo quantum memory protocol based on atomic frequency combs (AFC)[5]. It is rooted in the interaction of light with an ensemble of atomic absorbers, to date RE crystals cooled to cryogenic temperatures, with an inhomogeneously broadened absorption line that has been tailored into a series of equally spaced absorption peaks (see Fig. 2). The absorption of a single photon leads to a collective excitation shared by all $N$ atoms, and an atomic state of the form $|\Psi\rangle = \sum_{i=1}^{N} c_i e^{i2\pi\delta_i t} e^{-ikz} |g_1, \cdots e_i, \cdots, g_N\rangle$. The ground state of atom $i$ is denoted by $|g_i\rangle$, $\delta_i$ is the detuning of the atom's transition frequency from the light carrier frequency with wave vector $k$, $z$ is the position of the atom measured along the propagation direction of the light, and $c_i$ depends on its resonance frequency and position. Due to the particular shape of the tailored absorption line, the excited collective coherence rapidly dephases and repeatedly recovers after multiples of the *storage time* $T_s = 1/\Delta$, i.e. after a time that is pre-determined by the spacing of the teeth in the comb. This results in the re-emission of a photon in the state encoded into the original photon. Readout on demand can be achieved by temporarily mapping the optically excited coherence onto ground state coherence where the comb spacing is smaller, or the comb structure is washed out[5].

Our storage device, a Ti:Tm:LiNbO$_3$ optical waveguide cooled to 3 Kelvin, is detailed in Fig. 2. It was previously characterized in view of photon-echo quantum memory[28]. It combines interesting properties from the specific rare-earth element (795 nm storage wavelength), the host crystal (allowing for controlled de- and rephasing by means of electric fields), as well as from the wave-guiding structure (ease-of-use). We emphasize that lithium niobate waveguides have also been doped with erbium[7], and we conjecture that other rare earth ions can be used as well. This could allow extending the good properties of LiNbO$_3$ and the integrated approach to other storage wavelengths, ions with different level structures, etc.

To generate the AFC, we employ a sideband-chirping technique[29] (see methods) to transfer atomic population between different magnetic sublevels to create troughs and peaks in the inhomogeneously broadened absorption line. They form a 5 GHz wide comb with tooth spacing of 143 MHz, setting the storage time to 7 ns. Due to the specific level structure of thulium, the finesse of the comb in the broadband approach is two, which limits the memory efficiency to ~10%. However, experimental deficiencies decrease the efficiency in our experiment to ~2%. Together with fibre-to-waveguide in- and output coupling loss that we attribute to non-optimal mode overlap, this results in a system efficiency of ~0.2% (see supplementary information for a discussion of this limit).

To assess the quantum nature of our light-matter interface, we make projection measurements with the 795 nm photons and the 1532 nm photons onto time-bin qubit states characterized by Bloch vectors aligned along $\vec{a}, \vec{b}$, respectively, where $\vec{a}, \vec{b} \in [\pm\sigma_x, \pm\sigma_y, \pm\sigma_z]$ (see Fig. 3). Experimentally, this is done by means of suitably adjusted qubit analyzers, and by counting the number $C(\vec{a}, \vec{b})$ of detected photon pairs. The measurement is further explained in the methods. From two such spin-measurements, we calculate the normalized *joint-detection probability*

$$P(\vec{a}, \vec{b}) = \frac{C(\vec{a}, \vec{b})}{C(\vec{a}, \vec{b}) + C(\vec{a}, -\vec{b})}. \tag{2}$$

The results with the fibre delay line, and the memory, are detailed in table 1. From this data, we reconstruct the bi-photon states in terms of their density matrices $\rho_{in}$ and $\rho_{out}$, depicted in Fig. 3, using a maximum likelihood estimation[30]. Comparing $\rho_{in}$ and $\rho_{out}$, we quantitatively assess if the storage process has preserved entanglement by examining three standard figures of merit[31]: concurrence, entanglement of formation, and maximum possible S-parameter of the Clauser-Horne-Shimony-Holt (CHSH) Bell inequality[8] (see methods). To confirm entanglement, the first two figures must exceed zero (they are upper-bounded by one), while the latter indicates non-local quantum correlations if it exceeds two (it is upper-bounded by $2\sqrt{2}$). The results, listed in table 2, clearly show the presence of entanglement in $\rho_{in}$ and $\rho_{out}$ and, within experimental uncertainty, establish that the storage process preserves entanglement without measurable degradation.

Another condition for a perfect memory is that the difference between $\rho_{in}$ and $\rho_{out}$ vanishes. In other words, the memory should apply no unitary transformation to the stored qubit, resulting in the fidelity between $\rho_{in}$ and $\rho_{out}$ to be one. As indicated in table 2, and explained in the methods, we find $F=0.95\pm0.03$, showing that the quantum information originally encoded into the 795 nm photon is indeed faithfully retrieved. We point out that the original state (and hence the recalled state) has limited purity and fidelity with $|\phi^+\rangle$. This is due to the probabilistic nature of our SPDC source, which features a non-negligible probability to generate more than two photons simultaneously[27].

Our investigation provides an example of entanglement being transferred between physical systems of different nature, thereby adding evidence that this fundamental quantum property is not as fragile as is often believed. Furthermore, our broadband integrated approach allows linking a promising quantum storage device with extensively employed, high-performance sources of photons in bi- and multi-partite entangled states[8]. This opens the way to new investigations of fundamental and applied aspects of quantum physics. Having increased the storage bandwidth also significantly facilitates building future quantum networks, as mutual frequency matching of photons and distant quantum memories will be trivial. In addition, a large storage bandwidth, i.e. the possibility to encode quantum information into short optical pulses, allows increasing the number of temporal modes that can be stored during a given time. This enhances the flow of quantum information through a network and decreases the time needed to establish entanglement over a large distance using a quantum repeater[1,2].

We note that, parallel to this work, Clausen *et al.* have also demonstrated the storage and retrieval of an entangled photon using a neodymium-doped crystal[32].

**Acknowledgements** This work is supported by NSERC, QuantumWorks, General Dynamics Canada, iCORE (now part of Alberta Innovates), CFI, AAET and FQRNT. We thank Cecilia La Mela, Thierry Chanelière, Terence Stuart, Vladimir Kiselyov and Chris Dascollas for help during various stages of the experiment, Christoph Simon, Krishna Rupavatharam and Nicolas Gisin for discussions, and Alex Lvovsky for lending a single-photon detector.

**Author Information** The authors declare that they have no competing financial interests. Correspondence and requests for materials should be addressed to W.T. (wtittel@ucalgary.ca).

## Table 1| Joint-detection probabilities $P_{in}$ and $P_{out}$

|  | $\sigma_x \otimes \sigma_x$ | $\sigma_x \otimes \sigma_y$ | $\sigma_x \otimes \sigma_z$ | $\sigma_x \otimes$-$\sigma_z$ | $\sigma_y \otimes \sigma_x$ | $\sigma_y \otimes \sigma_y$ | $\sigma_y \otimes \sigma_z$ | $\sigma_y \otimes$-$\sigma_z$ |
|---|---|---|---|---|---|---|---|---|
| $P_{in}$ [%] | 90±2 | 49±1 | 49±1 | 51±1 | 52±1 | 10±2 | 51±1 | 49±1 |
| $P_{out}$ [%] | 89±6 | 49±8 | 48±4 | 52±4 | 49±6 | 14±5 | 49±4 | 51±4 |

|  | $\sigma_z \otimes \sigma_x$ | $\sigma_z \otimes \sigma_y$ | $\sigma_z \otimes \sigma_z$ | $\sigma_z \otimes$-$\sigma_z$ | -$\sigma_z \otimes \sigma_x$ | -$\sigma_z \otimes \sigma_y$ | -$\sigma_z \otimes \sigma_z$ | -$\sigma_z \otimes$-$\sigma_z$ |
|---|---|---|---|---|---|---|---|---|
| $P_{in}$ [%] | 46±1 | 46±1 | 94.2±0.1 | 5.8±0.1 | 46±1 | 45±1 | 7.6±0.2 | 93.0±0.2 |
| $P_{out}$ [%] | 51±6 | 56±6 | 94±1 | 6±1 | 48±5 | 52±5 | 6±1 | 94±1 |

Measured joint-detection probabilities for all projection measurements required to calculate the density matrices for the bi-photon state emitted from the source ($P_{in}$), and after storage and recall of the 795 nm photon ($P_{out}$). Uncertainties indicate one-sigma standard deviations.

## Table 2| Purity, entanglement measures and fidelities

|  | Purity [%] | Concurrence [%] | Entanglement of formation [%] | $S_{Max}$ | Fidelity with $|\phi^+\rangle\langle\phi^+|$ [%] | Input-Output Fidelity [%] |
|---|---|---|---|---|---|---|
| $\rho_{in}$ | 75.7±2.4 | 74.1±3.3 | 64.4±4.2 | 2.49±0.04 | 86.2±1.5 | 95.4±2.9 |
| $\rho_{out}$ | 76.3±5.9 | 74.5±8.3 | 65±11 | 2.49±0.10 | 86.6±3.9 | |

Purity, concurrence, entanglement of formation (normalized with respect to the entanglement of formation of $|\phi^+\rangle$), maximum possible S-parameter, and fidelity with the ideal state $|\phi^+\rangle$ for input and output density matrices $\rho_{in}$ and $\rho_{out}$. The input-output fidelity refers to the fidelity of $\rho_{out}$ with respect to $\rho_{in}$. Uncertainties indicate one-sigma standard deviations and are estimated using Monte Carlo simulation.

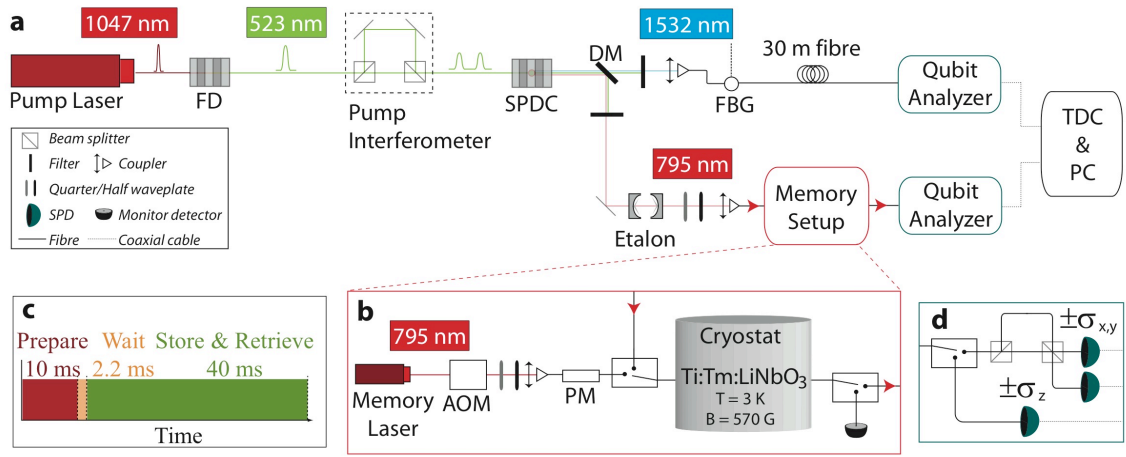

**Figure 1| Schematics of the experimental setup. a,** Generating and measuring entanglement. The *Pump Laser* generates 6 ps-long pulses at 1047.328 nm wavelength with an 80 MHz repetition rate that are frequency doubled (*FD*) in a $\chi^{(2)}$ non-linear, periodically poled lithium niobate (*PPLN*) crystal, resulting in 16 ps-long pulses at 523.664 nm wavelength with average power of 90 mW. Each pulse is coherently split into two by the unbalanced *Pump Interferometer,* featuring a 1.4 ns travel-time difference and phase locked using a stable reference laser. The pulses then undergo spontaneous parametric down-conversion (*SPDC*) in a second PPLN crystal, resulting in bright entangled twin-beams centred at $\lambda_1$=795.506 nm and $\lambda_2$=1532.426 nm. The two beams are separated on a dichroic mirror (*DM*) and, after removal of the pump light, frequency filtered using an *Etalon* and a fibre Bragg grating (*FBG*) with bandwidths of 6 GHz and 9 GHz, respectively. For sufficiently small pump powers, the non-vacuum part is well approximated by a single photon pair in a maximally entangled state. The 1532 nm photon is sent through a 30 m long standard telecommunication fibre and the 795 nm photon is either stored in the memory (setup described below) or sent through a fibre delay line (not pictured). To characterize the bi-photon state, we use *Qubit Analyzers.* Detection events are collected with a time-to-digital converter connected to a PC for analysis. **b,** Memory setup. The 795.506 nm continuous wave *Memory laser* passes through an acousto-optic modulator (*AOM*) and a 20 Gbps lithium niobate phase modulator (*PM*) that are used to intensity and phase/frequency modulate the beam as required for optical pumping of the Ti:Tm:LiNbO$_3$ waveguide. The waveguide is cooled to 3 K using a pulse tube cooler (*Cryostat*) and exposed to a 570 G magnetic field aligned with the crystal's C3-axis. Waveplates before the cryostat are used to adjust the polarization of the beam to the waveguide's TM mode. At the input and output of the cryostat the optical pump beam and the 795 nm photons are combined and separated by optical switches. **c,** Timing sequence. The timing sequence consists of three, continuously repeated phases: 10 ms of optical pumping (*Prepare*) as detailed above, 2.2 ms *Wait* time, which is set by the radiative lifetime of the $^3H_4$ state (see methods), and 40 ms *Store & Retrieve* time during which many 795 nm photons are successively stored in the waveguide and recalled after 7 ns. **d,** Qubit analyzer. The *Qubit Analyzers* consist of unbalanced interferometers featuring the same travel-time difference as the pump interferometer. They are phase locked to stable reference lasers and allow for projective measurements onto $\sigma_x$ and $\sigma_y$. Alternatively, fibre delay lines are used, enabling projections onto $\sigma_z$. 795 nm photons are detected by a single photon detector (*SPD)* based on a silicon avalanche photodiode (APD), and 1532 nm photons are detected by SPDs based on InGaAs APDs. To reduce complexity, the diagram details only the analyzer for the 1532 nm photons.

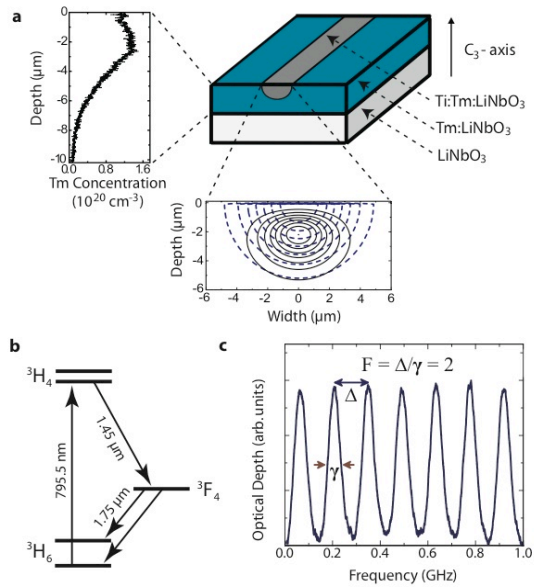

**Figure 2| The storage medium. a,** Waveguide geometry. The measured Tm$^{+3}$ concentration profile is given on the left and the calculated intensity distribution of the fundamental TM-mode at 795 nm wavelength below. Iso-intensity lines are plotted corresponding to 100%, 87.5%, 75% etc. of the maximum intensity. **b,** Simplified energy level diagram of Tm$^{+3}$ ions. The optical coherence time of the $^3H_6 \leftrightarrow {}^3H_4$ transition at 3 K is 1.6 μs, the radiative lifetimes of the $^3H_4$ and $^3F_4$ levels are 82 μs and 2.4 ms, respectively, and the branching ratio from the $^3H_4$ to the $^3F_4$ level is 44%. Upon application of a magnetic field, the ground and excited levels split into magnetic sublevels with lifetimes exceeding one second at 570 G magnetic field[28]. **c,** Atomic frequency comb. The bandwidth of our AFC is 5 GHz (here shown is 1 GHz broad section). The separation between the teeth, $\Delta$, is ~143 MHz, resulting in 7 ns storage time. The linewidth of the peaks, $\gamma$, is ~75 MHz, yielding a finesse $F$=2, as expected for the sinus-type comb.

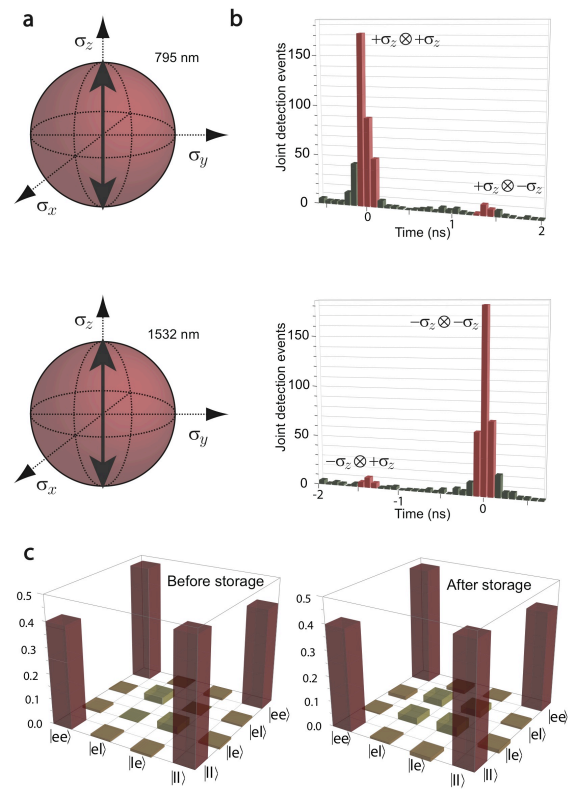

**Figure 3| Measurements and results. a,** Visualization of projection measurements. The measurement settings for the 795 nm (1532 nm) qubit analyzer are depicted on the upper (lower) Bloch sphere. The joint settings enable projections onto $\sigma_z \otimes \sigma_z$ and $\sigma_z \otimes$-$\sigma_z$. **b,** Results for joint projection measurement after storage. The top (bottom) histogram displays joint detection events for the projection onto $\sigma_z \otimes \sigma_z$ and $\sigma_z \otimes$-$\sigma_z$ (-$\sigma_z \otimes \sigma_z$ and -$\sigma_z \otimes$-$\sigma_z$) as a function of the time difference between detections of the 795 and the 1532 nm photons. The desired events are those within the red highlighted *time windows*. This allows calculating the joint-detection probability for this particular projection (for more results see table 1). **c,** Density matrices. Density matrices calculated using a maximum likelihood estimation for the bi-photon state before and after storage. Only the real parts are shown – the absolute values of all imaginary components are below 0.04.

**Methods**

**Calculation of purity, entanglement measures and fidelities.** From the density matrices ρ for the input and retrieved states we calculate the *purity*

$$P=tr(\rho^2). \tag{3}$$

Furthermore, assuming an arbitrary two-qubit input state ρ, *concurrence* is defined as

$$C(\rho)=\max\{0,\lambda_1-\lambda_2-\lambda_3-\lambda_4\}, \tag{4}$$

where the $\lambda_i$'s are, in decreasing order, the square roots of the eigenvalues of the matrix $\rho\sigma_y\otimes\sigma_y\rho^*\sigma_y\otimes\sigma_y$, and $\rho^*$ is the elementwise complex conjugate of ρ. The *entanglement of formation* is then calculated as

$$E_F(\rho) = H\left(0.5 + 0.5\sqrt{1 - C^2(\rho)}\right), \tag{5}$$

where H(x) = − x log₂ x − (1-x) log₂ (1-x). Finally, the maximum possible value of the CHSH-Bell inequality parameter is

$$S_{Max} = 2\sqrt{1 + C^2}, \tag{6}$$

and the fidelity between ρ and σ is

$$F(\rho, \sigma) = \left(tr\sqrt{\sqrt{\rho}\sigma\sqrt{\rho}}\right)^2. \tag{7}$$

**Preparation of the AFC.** The AFC amounts to a periodic modulation in frequency of the optical density of the inhomogeneously broadened $^3H_6 \leftrightarrow ^3H_4$ thulium absorption line. It can be generated by optically pumping atoms to off-resonant shelving levels - in our case nuclear Zeeman levels[28]. To that end, we modulate the intensity of the 795 nm memory laser while scanning its frequency[29]. The frequency sweep is implemented using a lithium niobate phase modulator driven by a 20 Gs/s arbitrary waveform generator. To avoid overlap of first and higher order modulation, the sweep extends from 5 GHz to 10 GHz, thus efficiently preparing a 5 GHz-bandwidth AFC memory. The laser intensity modulation is achieved by beating two frequency components, generated in an acousto-optic modulator (AOM) placed before the phase-modulator.

The memory storage time $T_S$ is set by the frequency spacing between the teeth of the AFC, and is determined by $T_S = \delta/\alpha$, where δ=0.35 MHz is the difference between the two frequency components and α=50x10⁶ MHz/s is the sweep rate. This yields 142.85 MHz spacing between the AFC teeth, which translates into 7 ns memory storage time. For a high contrast AFC, the chirp cycle is repeated 100 times leading to a 10 ms overall optical pumping duration. The 2.2 ms wait time following the preparation corresponds to 27 times the radiative lifetime of the $^3H_4$ excited level, and ensures no fluorescence masks the retrieved photons.

The optical pumping involves population transfer between ground-state sublevels that arise under the application of a magnetic field (nuclear Zeeman levels and levels due to super-hyperfine splitting)[28]. As the comb structure extends over all these levels, we carefully chose the magnetic field to make sure that those ions that initially absorb at frequencies where we desire a trough are transferred to frequencies where we desire a peak.

**The measurement.** First, we stabilize the pump interferometer and the 1532 nm interferometer to arbitrarily chosen phase values. We define the phase introduced by the pump interferometer to be zero, i.e. we absorb it into the definition of the *early* and *late* qubit states, as reflected by Eq. 1. Furthermore, we define the measurement performed by the 1532 nm qubit analyzer to be $+\sigma_x$. Next, we change the phase of the 795 nm interferometer and maximize the joint-detection probability without memory. We define this setting to correspond to a projection onto $+\sigma_x$. We measure the number of joint detection events over 5 minutes, and calculate the joint-detection probability $P_{in}(\sigma_x \otimes \sigma_x)$ according to Eq. 2. Next, we add the memory and similarly measure $P_{out}(\sigma_x \otimes \sigma_x)$ over approx. 5 hours. When necessary to change the setting of either qubit analyzer to $\sigma_y$, we increase the phase difference introduced by the respective interferometer by $\pi/2$. For projection measurements onto $\sigma_z$, we use the delay line. Each joint projection measurement is done with and without memory; the results are given in table 1.

## Supplementary information

**The Ti:Tm:LiNbO$_3$ waveguide.** To fabricate the Ti:Tm:LiNbO$_3$ waveguide, a commercially available 0.5 mm thick Z-cut wafer of undoped, optical grade congruent lithium niobate (CLN) was cut into samples of 12 mm x 30 mm size. Tm doping was achieved by in-diffusing a vacuum-deposited (electron-beam evaporated) Tm layer of 19.6 nm thickness. The diffusion was performed at 1130°C during 150 h in an argon-atmosphere followed by a post treatment in oxygen (1 h) to get a full re-oxidization of the crystal. Tm occupies regular Li-sites when incorporated in CLN by diffusion[1]. The Tm in-diffusion leads to a 1/e-penetration depth of about 6.5 μm. The maximum Tm concentration of about $1.35 \times 10^{20}$ cm$^{-3}$ corresponds to a concentration of 0.74 mole %, which is considerably below the solid solubility of Tm in CLN[2]. Subsequently, the waveguide was formed by the well-known Ti-indiffusion technique. At first, a 40 nm thick titanium layer was electron-beam deposited on the Tm-doped surface of the CLN substrate. From this layer, 3.0 μm wide Ti stripes were defined by photo-lithography and chemical etching and subsequently in-diffused at 1060°C for 5 h to form 30 mm long optical strip waveguides. In the wavelength range around 795 nm, the waveguides are single mode for TE- and TM-polarization. To finish the fabrication, the waveguide was cut to 15.7 mm and end faces were carefully polished normal to the waveguide axis.

**Limitation to efficiency.** The system efficiency of our implementation is currently limited to around 0.2%. While this is sufficient to show the entanglement-preserving nature of the storage process, it is clear that this number has to be improved to make the memory more practical and to allow for more involved fundamental measurements.

First, we note that better optical mode matching between the fibre and the LiNbO$_3$ waveguide can be expected to improve the fibre-to-fibre transmission from 10% to 50%.

Second, assuming storage in optical coherence and Gaussian-shaped teeth, the efficiency of the first recall in the forward direction is given by

$$\varepsilon = \left(d_1/F\right)^2 e^{-d_1/F} e^{-7/F^2} e^{-d_0},$$

(1)

where F denotes the finesse of the comb, and $d_1/F$ and $d_0$ are the reversible and irreversible optical depth[3]. In our case, the optical pumping required to generate the AFC is based on population transfer between ground-state sublevels that arise under the application of a magnetic field[4]. As the comb structure extends over all these levels, we carefully chose the magnetic field so that ions that absorb initially at frequencies where we desire a trough are shifted to frequencies where we desire a peak. However, this fixes the fidelity of the comb to two, as ions can only be 'shuffled around' but not removed from the spectral region covered by the SPDC photons. This impacts on the storage efficiency and sets an upper bound of ~10%. Yet, we note that the efficiency can be increased to ~17% when applying a phase-matching operation that results in backward emission of the stored photon[5]. Further improvement is expected when changing the teeth shape from Gaussian to square[6]. All options combined, it seems possible to achieve a system efficiency of around 15%, which is 75 times larger than in the current implementation.

We point out that the limitation due to the comb finesse is not necessarily a consequence of generating broadband combs, but of the relatively small Zeeman shifts of the thulium ground levels. For other ions, it may be possible to find atomic levels with increased energy spacing or greater sensitivity to magnetic fields[7], allowing for a larger finesse, hence larger efficiency.

**Longer storage time and on-demand readout.** Currently, the maximum storage time of our memory is approximately 300 ns. This value is determined by the minimum tooth spacing of the AFC, which is limited by spectral diffusion[4]. However, spectroscopic investigation of a $Tm:LiNbO_3$ bulk crystal may suggest that spectral diffusion decreases when lowering the temperature, similar to the observed improvement of the optical coherence time[8]. This implies the possibility to extend the storage time.

In addition, it may be possible to further improve the storage time and achieve on-demand recall by temporarily transferring the optically excited coherence to coherence between the $^3H_6$ and $^3F_4$ electronic levels, similar to storage of coherence in spin-waves[9]. However, the coherence properties and the suitability of the $^3F_4$ state for such a transfer remains to be investigated.